\documentclass[ twocolumn, showpacs, pra, superscriptaddress, floatfix
 amsmath,amssymb,
 aps,
floatfix,
]{revtex4-1}
\usepackage{CJK}

\usepackage{graphicx}
\usepackage{dcolumn}
\usepackage{comment}
\usepackage{xcolor}
\usepackage[colorinlistoftodos]{todonotes}
\usepackage{hyperref}
\usepackage[english]{babel}
\usepackage{bm}
\usepackage{amsmath}
\usepackage{amssymb}
\usepackage{comment}
\usepackage{subfigure}
\usepackage{epstopdf}
\usepackage{hyperref}
\usepackage{appendix}
\usepackage{float}

\usepackage{mathtools}

\DeclarePairedDelimiterX\braket[2]{\langle}{\rangle}{#1 \delimsize\vert #2}

\DeclareUnicodeCharacter{03B2}{\ensuremath{\beta}}
\UseRawInputEncoding

\makeatother

\newcommand{\Rnum}[1]{\uppercase\expandafter{\romannumeral #1\relax}}
\newcommand{\txrd}[1]{\textcolor{red}{#1}}

\begin{document}


\title{STIRAP simulations from the 1S to the 2S states for Hydrogen/Antihydrogen atoms}

\author{AbdAlGhaffar K. Amer}
 \email{amer1@purdue.edu}
 \affiliation{
 Department of Physics and Astronomy, Purdue University, West Lafayette, Indiana 47906 USA
}

\author{A. Capra}
 \affiliation{
 TRIUMF, 4004 Wesbrook Mall, Vancouver, British Columbia, Canada V6T 2A3
}

\author{T. Friesen}
\affiliation{
 Department of Physics and Astronomy, University of Calgary, Calgary, Alberta, Canada T2N 1N4
}

\author{M. C. Fujiwara}
 \affiliation{
 TRIUMF, 4004 Wesbrook Mall, Vancouver, British Columbia, Canada V6T 2A3
}

\author{T. Momose}
 \affiliation{
 TRIUMF, 4004 Wesbrook Mall, Vancouver, British Columbia, Canada V6T 2A3
}
\affiliation{
 Department of Chemistry, University of British Columbia, Vancouver, British Columbia, Canada V6T 1Z1
}

\author{C. So}
 \affiliation{
 TRIUMF, 4004 Wesbrook Mall, Vancouver, British Columbia, Canada V6T 2A3
}
\affiliation{
 Department of Physics and Astronomy, University of Calgary, Calgary, Alberta, Canada T2N 1N4
}

\author{F. Robicheaux}%
 \email{robichf@purdue.edu}
\affiliation{
 Department of Physics and Astronomy, Purdue University, West Lafayette, Indiana 47906 USA
}

\begin{abstract}
Achieving a high population of antihydrogen/hydrogen atoms in the 2S level is essential for spectroscopy measurements testing similarities between matter and antimatter. We propose and examine the efficiency of applying the STIRAP (Stimulated Raman Adiabatic Passage) process in achieving high population transfer from the 1S to the 2S levels. We utilized a circularly polarized Lyman alpha, Ly-$\alpha$, pulse to couple the 1S state to the 2P state and a microwave pulse to couple the 2P state and the 2S state. We calculate the efficiency of the STIRAP process for transferring the population between the stretched states $(1S_d, 2S_d)$ as a function of experimental parameters such as Rabi frequencies and pulse durations. We find that a Ly-$\alpha$ pulse with an energy of a few nanojoules could produce nearly perfect transfer at zero detunings for atoms on the laser beam axis. We extended the analysis to a thermal ensemble of atoms, where Doppler detuning affects the velocity distribution of the hydrogen atoms produced in the 2S level. We found that the width of such velocity distribution is controlled by the Rabi frequency. We show that the peak velocity of the hydrogen atoms in the 2S level after STIRAP can be controlled by the Ly-$\alpha$ pulse detuning. The efficiency of STIRAP in transferring population increases at low temperature (T$\sim$1~mK). Finally, we show that a background magnetic field improves the transfer rates between the other trappable states $(1S_c, 2S_c)$.
\end{abstract}


\maketitle

\section{\label{sec:level1}Introduction}

The 2S level of the hydrogen atom is a metastable state with a lifetime of 0.12 seconds, which is about 8 orders of magnitude longer than the 2P state \cite{parthey2011improved}. The 1S-2S transition is a good candidate for precision measurement experiments\cite{niering2000measurement}. For instance, it is used in testing the CPT(charge conjugation, parity, and time reversal) symmetry of the standard model \cite{baker2025precision}. These tests are achieved by comparing the different fine and hyperfine splittings of the hydrogen and antihydrogen atoms \cite{ahmadi2017observ,ahmadi2017observation, ahmadi2018characterization}. Besides increasing the accuracy of the 1S-2S transition frequency, future experiments to evaluate the antihydrogen Rydberg constant as well as the antiproton radius require two independent transition frequency measurements. A good candidate for such measurements are transitions from the 2S level to higher levels of the antihdrogen atoms. These measurements would benefit from a large population of antihydrogen atoms in the 2S level. One of the aims of this work is to show that the STIRAP process could produce useful populations of atoms in the 2S level with lower pulses' intensities and thus negligible ionization rates. 

Typically, the 1S-2S transition of the hydrogen atom is achieved by using 2 counter-propagating photons of wavelength 243 nm each \cite{parthey2011improved,matveev2013precision}. This approach removes the net first order Doppler shift, thus it mostly eliminates the Doppler broadening effects and produces a narrow transition linewidth. Additionally, the hydrogen atoms don't experience recoil when absorbing the two counter propagating photons. Therefore, this scheme is useful for spectroscopy measurements of the 1S-2S transition. However, when the aim is to produce a large population of hydrogen atoms in the 2S level, it can face some challenges. For instance, the 1S-2S transition is highly nonresonant in the first 243~nm photon of this scheme. Thus, the transition requires high laser intensities and energies \cite{parthey2011improved} \cite{ahmadi2017observation} reaching orders of magnitude $10^7$~$W\,m^{-2}$ and $10^{-3}$~J, respectively, for meaningful transition probabilities. At such high intensities, the rate of photo-ionization from the 2S state becomes high \cite{rasmussen2017aspects}, thus reducing the efficiency of the 1S-2S transfer process. The current efficiency of such a scheme in the literature is about 0.1\% \cite{rasmussen2017aspects}. However, as far as we know, that technique has not been optimized to produce maximum population transfer. Other studies have investigated population transfer through the continuum \cite{carroll1992coherent,nakajima1994population,yatsenko1997population}. However, in \cite{yatsenko1997population} it was shown that when such technique is applied to hydrogen, the ionization rates are comparable to those seen in the 243~nm photons scheme.




In this work, we simulate the implementation of a STIRAP process to transfer hydrogen or antihydrogen atoms from the 1S to the 2S level. We use the 2P level as the intermediate state, relying on the resonant dipole transitions between the 2P levels and both the initial 1S and the target 2S levels. Two radiation pulses are utilized; the Probe, P, pulse, which is a Lyman alpha pulse, Ly-$\alpha$, that couples the 1S and the 2P levels and a Stokes, S, pulse which is a microwave pulse that couples the 2P and the 2S levels. The required Ly-$\alpha$ pulse has intensities of $\sim 10^4$ W m$^{-2}$ and energies of $\sim 10^{-9}$ J, several orders of magnitude less than those used in the 243~nm two-photon excitation. Consequently, the ionization probability becomes negligible (of order $10^{-8}$ of the excited atoms). A possible limitation to applying STIRAP to this system is the high decay rate from the intermediate 2P level to the ground state compared to the almost negligible decay rate into the 2S level. This high decay rate as well as the disparity in the branching ratios are known to reduce the efficiency of the STIRAP process \cite{ivanov2005spontaneous}. Another possible limitation of applying this approach 
is that the Doppler shift experienced by the Ly-$\alpha$ pulse is five orders of magnitude higher than that experienced by the microwave pulse connecting the 2S and the 2P levels \cite{kramida2010critical}. For thermal ensembles of hydrogen atoms at millikelvin temperatures, this shift results in single and two-photon detunings comparable to transition widths of interest. This affects the efficiency of the STIRAP process in transferring populations into the 2S level for higher temperature experiments. On the plus side, the Doppler effect can also be used to generate a population of atoms in the 2S level with a specific range of velocities, Sec. \ref{subsec.thermal_distrib}. Previous studies have been conducted on the velocity distribution of atomic systems after STIRAP for cooling purposes \cite{ivanov2012theory,ivanov2012robust,vovk2019features}. 


In this paper, we show the feasibility of using STIRAP to transfer population to the 2S hydrogen atom level. In the methods section, Sec. \ref{sec_methods}, we introduce the setup of the STIRAP process for a single hydrogen atom. We start with presenting the Lindbladian for the STIRAP process between the stretched states $(1S_d, 2S_d)$ in Sec. \ref{subsec:full treatment}. Then, in Sec. \ref{subsec.mom_rec} we discuss the approximate method we implement to obtain the velocity distribution of the atoms after the STIRAP process, including the recoil effects from absorbing the Ly-$\alpha$ photon as well as the decay from the intermediate state. Later, in Sec. \ref{subsec.B_methods} we discuss how the presence of a magnetic field affects the STIRAP Hamiltonian. Sec. \ref{sec_results} is the results section where we discuss the efficiency of applying the STIRAP approach in producing an appreciable transfer of population from the 1S level to the 2S level given the properties of the atom cloud and the geometric shape of the STIRAP beams. We divide it into three parts. 
We start in Sec. \ref{subsec_single_STIRAP} by analyzing the features of the single atomic system with emphasis on the ones relevant to understanding a bulk of atoms behavior.
Sec. \ref{subsec_thermal_ens} discusses the velocity distribution of the hydrogen atoms in both the 1S and the 2S levels after the STIRAP process. In Sec. \ref{subsec_res_inc_B} we show that the presence of a magnetic field can improve the STIRAP process between the non stretched states $(1S_c, 2S_c)$ with relatively little loss in the hydrogen atoms. We will use ``hydrogen" atom in all that follows to refer to both hydrogen and antihydrogen with the understanding that for antihydrogen, the spins are flipped to obtain the states with the appropriate properties in a magnetic field. 

\section{\label{sec_methods}Methods}

 \subsection{\label{subsec:full treatment} Treatment of STIRAP between the 1S and 2S states of hydrogen}

 To transfer population from the 1S to the 2S level using STIRAP, we couple these levels through the intermediate 2P states using two pulses; a Probe pulse and a Stokes pulse \cite{RevModPhys.89.015006}. 
 A $\Lambda$ linkage is constructed using the 2P states with a fine structure total angular momentum (J=3/2). The 2P states with fine structure total angular momentum (J = 1/2) are detuned enough that the coupling to them is negligible; we tested this by numerical calculations in addition to simple estimates from Rabi frequencies and detunings. 
 
 \begin{figure}[h!] 
  \includegraphics[width=1.0\linewidth]{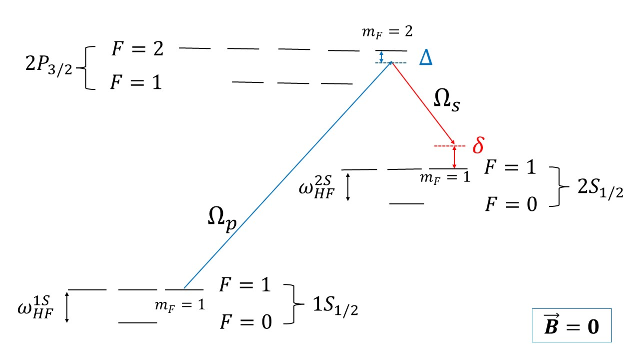} 
\caption{\label{Energy levels} An illustration of the transitions of interest with the 1S and 2S hyperfine splitting; $\omega_{HF}^{1S}= 2\pi \times 1.42\times10^9$~rad/s, $\omega_{HF}^{2S}=2\pi \times 0.177\times10^9$~rad/s \cite{kramida2010critical,diermaier2017beam} not drawn to scale. The 2P hyperfine splitting is much smaller (23~MHz) and also not drawn to scale. The pump pulse is the Ly-$\alpha$ pulse with frequency $\omega_{\alpha} = 2 \pi \times 2.466 \times 10^{15}$ rad/s, and the Stokes pulse is a microwave pulse with frequency $\omega_s = 2 \pi \times 9.78 \times 10^{9}$~rad/s. In the $\Lambda$ scheme $\Delta = \delta_p$ and $\delta = \delta_p - \delta_s$, where $\delta_p$ and $\delta_s$ are the P pulse and the S pulse detunings respectively, with $\delta_p$ as defined in Eq. (\ref{Pulse_field}). The 2P states with electron angular momentum $J=\frac{1}{2}$ are detuned by about $11$~GHz so their contributions are negligible.} 
\end{figure}
 The Probe, P, pulse is the Lyman alpha pulse, Ly-$\alpha$, that couples the 1S triplet (F=1) and the 2P fine structure state with the electron total angular momentum (J = 3/2) as illustrated in Fig. \ref{Energy levels}. In our simulations, we take the P pulse to be circularly polarized with a Gaussian profile in time, such that:
\begin{equation}
\label{Pulse_field}
        \vec{E}_p(t) = - \mathfrak{Re}\left(E^0_p e^{ - i \left(\omega_{\alpha} +  \delta_p \right) t} e^{-(2 \ln{2}) (t-t_0^p)^2 /\tau^2}\hat{\epsilon}_+ \right)
\end{equation}
with the unit vector $\hat{\epsilon}_+ = \frac{-1}{\sqrt{2}}(\hat{x}+i \hat{y})$, $E^0_p$ is the amplitude of the electric field, the Ly-$\alpha$ transition angular frequency $\omega_{\alpha} = 2 \pi \times 2.466 \times 10^{15}$ rad/s \cite{kramida2010critical}, $\delta_p$ is the detuning in the angular frequency, the FWHM of the intensity duration, $\tau$, was taken to be $20$~ns in all our simulations to be similar to the expected time scale in the proposed HAICU experiments. The pulse also has a Gaussian radial profile with the maximum amplitude of the electric field $E^0_p$ given by: 
\begin{equation}
    E^0_p = \sqrt{\frac{\mathcal{E}}{ w_0^2 \tau}} \sqrt{\frac{4 (2 \sqrt{\ln{2}})}{\epsilon_0 c \pi^{3/2}}}
\end{equation}
where, $w_0$ is the waist of the Gaussian pulse, $\mathcal{E}$ is the energy in the pulse, $c$ is the speed of light, and $\epsilon_0$ is the permittivity of free space. In many simulations, we used a pulse waist 
$w_0 =1$~mm and a total energy of $3$~nJ. We made such choices for the pulse parameters to match those proposed in experiments on trapped hydrogen and antihydrogen atoms for the HAICU project \cite{fujiwara2020demonstration,capra2024observation,fujiwara2023fundamental} at TRIUMF (Vancouver, Canada). This pulse produces a Hamiltonian matrix element, as shown in App.\ref{app.eval_Hammat}, of the form: 
\begin{eqnarray}
\label{eq.dip_mat}
        H_{1 2}(t) &=& -\hbar \Omega_p(t)/2 = \frac{1}{2}
        \left< \psi_{1 0 0}|e \vec{E}_p(t) \cdot \vec{r}|\psi_{2 1 1} \right> \nonumber\\ 
         &=& -\frac{e E^0_p}{2\sqrt{3}} \left< R_{10} (r)|r|R_{21}(r) \right> e^{-(2 \ln{2}) (t-t_0^p)^2 /\tau^2} \nonumber \\
         &=& -\frac{\hbar \Omega^0_p}{2} e^{-(2 \ln{2}) (t-t_0^p)^2 /\tau^2}
\end{eqnarray}
where $t_0^p$ is the time at which the Gaussian pulse reaches its peak and $R_{n l}$ is the radial part of the hydrogen eigenstates $\psi_{n l m}$. This gives a peak Rabi frequency of the P pulse $\Omega^0_p = 4.915\times10^8$~rad/s. When discussing ensembles of trapped atoms in Sec. \ref{subsec.thermal_distrib} we will average over the atom's position within the radial profile of the pulse.

 The Stokes, S, pulse is a microwave pulse that couples the 2P states (J = 3/2) with the 2S hyperfine states and has an angular frequency $\omega_s = 0.978 \times 2 \pi \times 10^{10}$~rad/s and a detuning $\delta_s$. We used the same electric-field polarization for the S pulse as that for the P pulse, Eq. (\ref{Pulse_field}) but with a different amplitude $E^0_s$ instead of the amplitude of the P pulse $E^0_p$. We also use a Gaussian time dependence for the S pulse which gives a Hamiltonian matrix element of the form: 
\begin{eqnarray}
\label{eq.s_pulse}
         H_{23}(t) &=& -\hbar \Omega_s(t)/2 \nonumber\\ 
         &=& -\frac{\hbar \Omega^0_s}{2} e^{-(2 \ln{2})t^2 /\sigma^2}
\end{eqnarray}
with $\Omega^0_s$ and $\sigma$ are the parameters controlling the strength and the duration of the S pulse respectively. In most of the simulations presented in this paper $\Omega^0_s$ is of the same order of magnitude as $\Omega^0_p$. As typically known for efficient STIRAP\cite{RevModPhys.89.015006},\cite{bergmann2015perspective} the parameters $\sigma$ and $t_0^p$ need to be tuned such that the Pump-Stokes offset \cite{bergmann2015perspective} creates a `counter-intuitive' ordering of the pulses with the S pulse preceding the P pulse and a sufficient overlap between them. As the parameters deviate from such optimum configuration, the efficiency of the STIRAP process in transferring population decreases, and more population moves into the radiative intermediate state. An example of the time dependence of the two pulses is shown in Fig. \ref{prob_omg_vs_t}. These pulses are more closely spaced in time than is typical for STIRAP, as will be discussed in Sec. \ref{subsec_single_STIRAP}. Unlike the P pulse, the microwave S pulse is not a Gaussian pulse in the radial direction; rather, we assume that its amplitude is approximately unchanged over the sample size of the hydrogen atoms. 

In the complete treatment of the STIRAP process, there are 16 possible states in the system; namely 4 hyperfine states for the 1S level, 4x2 states for the 2P (J = 3/2) level, and 4 hyperfine states for the 2S level as illustrated in Fig. \ref{Energy levels}. However, the circularly polarized pulses, Eq. (\ref{Pulse_field}), impose well-known selection rules for a rank one tensor operator \cite{georgi2000lie}. Due to these selection rules, only 5 of the 2P level states ($F$,$m_F$) couple to the S level states, namely the 2P states: $\{(2,2),(2,1),(2,0),(1,1),(1,0)\}$. Thus, resulting in only 13 of the 16 possible states being accessible through the circularly polarized pulses. Thus, the STIRAP Hamiltonian will be a matrix of dimensions $13\times13$.

In order to include the decay from the 2P states to the 1S states we use the Lindblad master equation for the density matrix of the system $\rho$: 
\begin{equation}
\label{master_eq}
    \dot{\rho} = \frac{-i}{\hbar} [H,\rho] - \sum_n  \Gamma_n \Big( \left(L_n^\dagger L_n \rho + \rho L_n^\dagger L_n\right)/2 - L_n \rho L_n^\dagger \Big)
\end{equation}
with H being the hermitian STIRAP Hamiltonian and the $\Gamma_n$ are the decay rates from the 2P to the 1S states for each decay channel. They are proportional to the square of the expectation value of the dipole matrix element connecting the initial and final states \cite{shankar2012principles}. The $L_n$ matrices are $13\times13$ dimensional jump operators. They transfer population from the 2P states to the relevant 1S states with the decay rate $\Gamma_n$ as defined in App.\ref{app.decay_rates}.

Generally, the STIRAP Hamiltonian couples each 1S hyperfine eigenstate to multiple 2P hyperfine eigenstates. By explicit evaluation of the Hamiltonian, one observes that the stretched state $1S_d$ ($F=1$ , $m_F=1$) only couples to the $2P_a^{\uparrow}$ ($F=2$ , $m_F=2$) which only couples through the S pulse to the states $2S_d$ ($F=1$ , $m_F=1$). Thus, those three states form a closed system under the Hamiltonian part of the time evolution with the effective $3\times3$ STIRAP Hamiltonian:

\begin{equation}
\label{full_Ham}
\mathcal{H}(t) = -\hbar \begin{pmatrix}
    0 & \frac{\Omega_p}{2} & 0 \\
    \frac{\Omega_p}{2} & \Delta & \frac{\Omega_s}{2}  \\
    0 & \frac{\Omega_s}{2} & \delta 
\end{pmatrix}(t)
\end{equation}
where $\{\Delta, \delta\}$ are the single-photon and the two-photon detuning, respectively, see caption of Fig. \ref{Energy levels}. Similarly, the $2P_a^{\uparrow}$ state can only decay to the $1S_d$, contrary to the other 2P states that could decay to multiple states of the 1S level. Consequently if the hydrogen atoms are initialized in the $1S_d$ state, then the time evolution of the density matrix under the full master equation doesn't move population outside the three states: $1S_d$, $2P_a^{\uparrow}$ and $2S_d$. Consequently, we can reduce the treatment of the problem to a three state subsystem constructed from these states. 

We work in this $3\times3$ subspace as we examine the efficiency of the STIRAP process and its dependence on the different experimental parameters. In Sec. \ref{subsec_res_inc_B} we show that a magnetic field can help selectively transfer population between the other trappable non-streched states; $1S_c$ and $2S_c$. That is achieved by making the subspace of interest approximately closed. Finally, we note that, in the absence of a magnetic field, the usage of laser pulses with the opposite circular polarization would only flip the total angular momentum F of the states involved in the STIRAP process. 

\subsection{\label{subsec.mom_rec} Thermal distribution and recoil effects}
In the previous section, the discussion did not include the center of mass motion of the hydrogen atom. 
In App.\ref{app.Ham_w_centerofmass} we show the change in the STIRAP Hamiltonian when we work with a tensor product of the internal states and the center of mass momentum state in the direction of the incident Ly-$\alpha$ photon $\left| k \right>$. In the literature, there has been work on the use of STIRAP processes in atomic systems for cooling \cite{ivanov2012theory,ivanov2012robust,vovk2019features}. Here, we present the numerical technique we used to study the velocity distributions of the atoms after the STIRAP process in both the initial and the target states. This atomic system gains an effective momentum kick from the STIRAP process as well as from the decay of the intermediate state, which are both accounted for as follows.

We restrict the analysis to the states: $\left\{1S_d \otimes \left|k\right>, 2P_a^{\uparrow} \otimes \left|k + k_\alpha\right>, 2S_d\otimes \left|k+ k_\alpha\right>\right\}$, where the first part is the internal state of the hydrogen atom and the second part is the center-of-mass momentum state in the direction of the incident photon. The states with the principle quantum number $n=2$ have gained a momentum kick $\hbar k_\alpha = 2 \pi \hbar/ (121.6\times10^{-9}~m)$ from absorbing the Ly-$\alpha$ photon. 


In appendix \ref{app.Ham_w_centerofmass}, we derive the Hamiltonian for a hydrogen atom with an arbitrary initial momentum $\hbar k$. The full density matrix of an atom in the thermal distribution would have different initial momenta for the 1S state, sampled from a Maxwell-Boltzmann distribution. 
In order for the density matrix to describe the entire system, we expand it to become $3M\times3M$, where $M$ is the number of the different allowed initial momentum $\hbar k$. We only keep track of the component of the momentum that is parallel to the incident P pulse. The momentum states have a grid spacing $\hbar \delta k$ in the interval $k \in [-k_{max}: k_{max}]$.
In the simulation $k_{max}$ is chosen based on the temperature of the initial hydrogen atoms such that all accessible momentum states are included. The momentum grid spacing $\hbar \delta k$ is chosen to be sufficiently smaller than the recoil momentum $\hbar k_\alpha$ as discussed in Sec. \ref{subsec_thermal_ens}.


The decay transfers population from the state $2P_a^{\uparrow} \otimes \left|k + k_\alpha\right>$ to the state $1S_d \otimes \left|k'\right>$, where $k'$ could take values in the range $[k:k+2k_\alpha]$. The exact value of momentum of the hydrogen atom after emitting a photon depends on the direction of the emitted photon compared to the atom momentum. A crude approximation would be to use the average outcome of multiple decay channels. Since hydrogen atoms are equally likely to emit a photon in the forward and the backward directions, the average effect would be no change in the momentum of hydrogen atoms after decaying to the 1S state, i.e $k'=k+k_\alpha$. We term this the ``averaged approach".


An improvement to this approximation is to allow the photon to be emitted in all directions, following the known angular distribution of dipole radiation $1 + \cos(\theta)^2$ \cite{milton2024classical}, with $\cos(\theta) = -(k'-k-k_\alpha)/k_\alpha$. Thus, the probability of the hydrogen atom 2P state to decay with momentum $\hbar(k+k_\alpha)$ into a 1S state with momentum $\hbar k'$ would be:
\begin{equation}
    \xi(k') = \frac{3}{8k_\alpha}\int_{k'-\delta k/2}^{k'+\delta k/2} \left(1+f(\bar{k})^2\right) d\bar{k}
\end{equation}
where $f(\bar{k})=\left(\bar{k}-k-k_\alpha\right)/{k_\alpha}$ and $\delta k$ is the momentum spacing of the density matrix in the simulation and $\bar{k} \in [k:k + 2k_\alpha]$. The simulations converged for $\delta k\simeq k_\alpha/20$. For later discussions, we term this approach with a quantized momentum grid the `quantized approach'. In Sec. \ref{subsec_thermal_ens} we compare the resulting thermal distributions from the crude `averaged approach' and the `quantized approach'. 

\subsection{\label{subsec.B_methods} Adding magnetic field effects}

The feasibility of using magnetic fields to selecticly control the transfer between magnetic sublevels has been explored earlier\cite{martin1995coherent,martin1996coherent,shore1995coherent}. For a hydrogen atom in the presence of a background magnetic field, the Hamiltonian of the system changes to:
\begin{equation}
    \hat{H} = \hat{H}_0 + \frac{g_e e}{2 m_e} \vec{B}\cdot\hat{\vec{S}}_e - \frac{g_p e}{2 m_p} \vec{B}\cdot\hat{\vec{S}}_p
\end{equation}
where $\hat{H}_0$ is the hydrogen atom Hamiltonian without the magnetic field, $e$ is the elementary charge, $m_e$ the electron mass, $m_p$ the proton mass, $\{\hat{\vec{S}}_e, \hat{\vec{S}}_p\}$ are the spin operators for the electron and the proton respectively and $\{g_e , g_p\}$ are the g-factor for the gyromagnetic ratios of the electron and the proton respectively. Due to the coupling to the magnetic field, the hyperfine states of the 1S and the 2S states are mixed. The eigenstates and their eigenvalues as a function of the magnetic field are found analytically using the Briet-Rabi formula. The stretched states discussed in the previous sections: $1S_d, 2P_a^{\uparrow}$ and $2S_d$, will remain a closed system with the magnetic field only introducing shifts in the transition frequencies. In this subsection we treat the evolution of the system starting from the trappable $1S_c$ state to the $2S_c$ state through the $2P_a^{\downarrow}$ state with (-1/2) proton spin. The states are given in terms of a tensor product between the orbital state $\left|n,l,m\right>$ and the z-component's spin states of the electron and the proton $\left|s^e_z,s^p_z\right>$ as follows:

\begin{eqnarray}
\label{eq.states_w_B}
    1S_c &=& \left|1,0,0\right> \otimes \big(\beta_{1S} \left|1/2, -1/2\right> + \alpha_{1S} \left|-1/2, 1/2\right>\big) \nonumber \\
    2P_a^{\downarrow} &=& \left|2,1,1\right> \otimes \left|1/2, -1/2\right> \\
    2S_c &=& \left|2,0,0\right> \otimes \big(\beta_{2S} \left|1/2, -1/2\right> + \alpha_{2S} \left|-1/2, 1/2\right>\big) \nonumber
\end{eqnarray}
with
\begin{eqnarray}
    \beta^2_n &=& \frac{1}{2} + \frac{B \mu_{+}}{\sqrt{4 B^2 \mu_{+}^2 + \hbar^2 \omega_{HF}^{n}}}\nonumber \\
    \alpha^2_n &=& \frac{1}{2} - \frac{B \mu_{+}}{\sqrt{4 B^2 \mu_{+}^2 + \hbar^2 \omega_{HF}^{n}}}
\end{eqnarray}
where $\omega_{HF}^{n}$ are the hyperfine splittings in Fig. \ref{Energy levels} with the index $n$ going over the 1S and the 2S levels and 
\begin{eqnarray}
    \mu_{+} = \frac{g_e e}{8m_e} + \frac{g_p e}{8m_p}
\end{eqnarray}
 The values of the $\{\alpha^2_n, \beta^2_n\}$ give the probability that the electron has a spin up or spin down respectively. Thus when evaluating the dipole matrix element in Eq. (\ref{eq.dip_mat}) for the STIRAP process, the matrix element coupling the $2P_a^{\downarrow}$ and the $1S_c$ states will have a factor of $\beta_{1S}$ while that coupling the $2P_a^{\downarrow}$ and the $2S_c$ states will have a $\beta_{2S}$ factor. 

The STIRAP Hamiltonian terms are thus dependent on the magnetic field:
\begin{equation}
    H(t) = \hbar \begin{pmatrix}
    E^{1S}(B) & -\frac{\Omega_p}{2} \beta_{1S} & 0 \\
    -\frac{\Omega_p}{2} \beta_{1S} & -\Delta + 2B \frac{g_e e}{\hbar m_e}& -\frac{\Omega_s}{2} \beta_{2S}  \\
    0 & -\frac{\Omega_s}{2} \beta_{2S} & -\delta + E^{2S}(B)
\end{pmatrix}(t)
\end{equation}
with 
\begin{equation}
    E^n(B) = -\frac{\omega_{HF}^{n}}{2} + \frac{\sqrt{4B^2 \mu_{+}^2 + \hbar^2 (\omega_{HF}^{n})^2}}{2\hbar}
\end{equation}
where again the index $n$ goes over the 1S and the 2S levels. The Hamiltonian could be put in a simpler form by subtracting $E^n(B)$ and redefining the detuning to become:
\begin{equation}
    H(t) = -\hbar \begin{pmatrix}
    0 & \frac{\Omega_p}{2} \beta_{1S} & 0 \\
    \frac{\Omega_p}{2} \beta_{1S} & \bar{\Delta}(B) & \frac{\Omega_s}{2} \beta_{2S}  \\
    0 & \frac{\Omega_s}{2} \beta_{2S} & \bar{\delta}(B)
\end{pmatrix}(t)
\end{equation}
where now
\begin{eqnarray}
\label{eq_B_det}
    \bar{\Delta}(B) &=& \Delta - 2B \frac{g_e e}{\hbar m_e} + E^{1S}(B) \nonumber \\
    \bar{\delta}(B) &=& \delta -  E^{2S}(B) + E^{1S}(B)
\end{eqnarray}

The other effect of the magnetic field is the change in the branching ratios of the 2P state when decaying into the 1S sublevels. Since the decay rate from $2P_a^{\downarrow}$ to $1S_c$ is proportional to the square of the dipole matrix element that connects them, the decay rate into $1S_c$ state becomes $\Gamma_0 \times \beta^2_{1S}$ where $\Gamma_0$ is as defined in App.\ref{app.decay_rates}. The decay into the $1S_a$ state has the branching ratio $(1-\beta^2_{1S})$. However, the $1S_a$ to the $2P_a^{\downarrow}$ transition is far from resonance so the $1S_a$ behaves as a dark state with negligible percentage of atoms getting back from it into the $\{1S_c, 2P_a^{\downarrow}, 2S_c\}$ subsystem.

Finally we note that in this section effects like the difference between the $g_e$ factor in the levels with the principle quantum numbers, n=1 and the n=2, as well as the diamagnetic potential term from the magnetic field were neglected, as both of them introduce energy changes of order of a few kilohertz \cite{rasmussen2017aspects} which result in much less than a percent change in the final populations; see Fig. \ref{prob_vs_detp_1}.



 
\section{\label{sec_results}Simulation Results}
Here we discuss the simulation results as we progressively add effects to the STIRAP process. In Sec. \ref{subsec_single_STIRAP}, we study the efficiency of the STIRAP process in transferring the population between the stretched states $\{1S_d, 2S_d\}$, when the atoms are held in space and no recoil effects are included. These two restrictions lead to cases where $\sim100\%$ of the population can be transferred to the 2S state. We vary physical parameters such as the detuning of each pulse and the Rabi frequency. In most of the simulations the STIRAP efficiency is limited by the realistic energy achievable by a Ly-alpha pulse in a setup like HAICU. 
In Sec. \ref{subsec_thermal_ens} we analyze the STIRAP process for thermal ensembles of hydrogen atoms; the resulting Doppler width for achievable temperatures can strongly reduce the fraction transferred into the 2S state. We also study how the change in the hydrogen atoms' temperatures affect the transfer rates of STIRAP as well as the velocity distribution of the resulting hydrogen atoms in the 2S level. In Sec. \ref{sec_ly-spatial} we study the dependence of the number of atoms going through STIRAP from a spatially extended sample on the P pulse's waist and energy with the spatial Gaussian profile of the Ly-$\alpha$ beam taken into account. Allowing the atoms to not be at the center of the Ly-$\alpha$ beam further reduces the population transferred into the 2S state and necessitates the introduction of a parameter analogous to a cross section per laser shot.
In Sec. \ref{subsec_res_inc_B} we treat the STIRAP between the states $\{1S_c, 2S_c\}$ with a background magnetic field that changes the efficiency of the transition due to the changing composition of the $1S_c$ state, Eq. (\ref{eq.states_w_B}).

\subsection{Stationary single-atom $1S_d \rightarrow 2S_d$ STIRAP}
\label{subsec_single_STIRAP}

Having made experimentally motivated choices for the parameters of the P pulse: $\Omega^0_p, \tau$ and the pulse waist $w_0$, there remain other experimental parameters that could be tuned to produce an efficient STIRAP process, namely $\{\Omega^0_s, \sigma, t_0^p\}$. To understand the effects of these parameters, we investigate the ideal case in which the atoms are fixed in space at the center of the Ly-$\alpha$ beam. In order to know what values for these parameters would produce an effective STIRAP process, we initialized the hydrogen atom in the $1S_d$ state and swept the values of those parameters over the ranges: $\Omega^0_s \in (0:5]\Omega^0_p$, $\sigma \in (0:10]\tau$ and $t_0^p \in (0:2]\sigma$. The optimum values for these parameters were found to be $\Omega^0_s = \Omega^0_p$, $\sigma=\tau$. The optimal pulse delay was found to be $t_0^p = 10.5 ~ ns \sim 0.525\tau$, which is smaller than $\sim0.85\tau$ which is typical in the literature \cite{vitanov1997analytic}. This difference is attributed to the high decay rate of the 2P level, App. \ref{app.decay_rates}, as well as the disparity in its branching ratios to the 1S states compared to the negligible decays into the 2S states. To test that, we ran separate simulations for smaller decay rates $\sim0.01\Gamma_0$ and equal branching ratios and both lead to more population transfer to the 2S level at larger pulses' separations when keeping the other pulses' parameters fixed. The case of equal branching ratios had better agreement with the $\sim0.85\tau$ in the literature \cite{RevModPhys.89.015006}. However, for the decay rates and the branching ratios of the 2P states of the hydrogen atom, App. \ref{app.decay_rates}, a pulses' separation of $t_0^p = 10.5 ~ ns \sim 0.525\tau$ gave the best population transfer into the 2S level. We examined implementing microwave pulses proportional to $e^{-t^4/\sigma^4}$ and obtained results similar to a Gaussian pulse. Unless otherwise indicated, we use the pulses' parameters listed in this paragraph for the rest of the simulations in this work.

\begin{figure}[h!] 
  \includegraphics[width=1.0\linewidth]{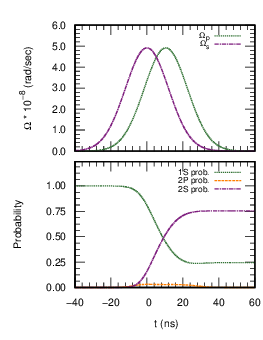} 
\caption{\label{prob_omg_vs_t} The S and P pulses as well as the states $\{1S_d, 2P_a^{\uparrow}, 2S_d\}$ probabilities as a function of time for typical experimental parameters and zero pulses detunings. The P pulse is the Gaussian in Eq. (\ref{eq.dip_mat}) with $t_0^p = 10.5$~ns. The S pulse follows Eq. (\ref{eq.s_pulse}) with $\sigma = \tau = 20$~ns and $\Omega^0_s = \Omega^0_p = 4.915\times10^8$~rad/sec. The maximum population in the 2P state is about 3\% which shows that the system is close to adiabatic.} 
\end{figure}

\begin{figure}[h!] 
  \includegraphics[width=1.0\linewidth]{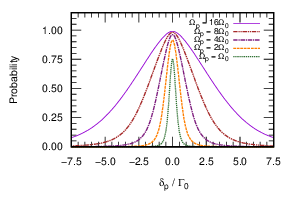} 
\caption{\label{prob_vs_detp_1} The final probability to be in the $2S_d$ state as a function of the P pulse detuning $\delta_p$ for the Rabi frequencies of the S and P pulses $\Omega^0_p = \Omega^0_s =\{1,2,4,8,16\}\times\Omega_0$, with $\Omega_0 = 4.915\times10^8$~rad/sec. The simulations were run while keeping the microwave pulse detuning $\delta_s$ as well as the pulses duration $\tau = \sigma = 20$~ns fixed; thus, changing the Ly-A pulse energy. The transfer rate to the 2S level drops for $\delta_p$ outside the range $[-\Omega^0_p:\Omega^0_p]$. } 
\end{figure}

\begin{figure}[h!] 
  \includegraphics[width=1.0\linewidth]{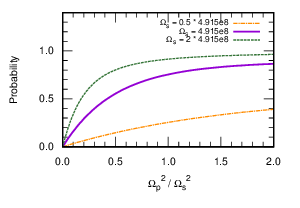} 
\caption{\label{Prob_vs_omgp} The transfer probability into the $2S_d$ state as $\Omega^0_p$ is varied while keeping $\Omega^0_s$ fixed at zero pulse detunings. The transfer probability is linear in $(\Omega^0_p)^2$ for small $(\Omega^0_p)^2/(\Omega^0_s)^2$.} 
\end{figure}

Since in our system $\Gamma_0 \gtrsim \Omega_0$, the analysis in \cite{ivanov2005spontaneous} works well in characterizing the adiabaticity domain of the system, where the STIRAP efficiency improves exponentially in $(\Omega^0_p)^2  \times \tau / \Gamma_0$. In Fig. \ref{prob_omg_vs_t} we show the S and P pulses of the STIRAP process as well as the time evolution of the probability of being in each of the three states as a function of time at zero detunings for both pulses. For these Rabi frequencies, the system falls into the adiabaticity domain where $\Omega^0_p \times \tau > 3\pi$ and $(\Omega^0_p)^2  \times \tau / \Gamma_0= 7.69 > 1$ \cite{RevModPhys.89.015006,ivanov2005spontaneous}.
The STIRAP process transfer rate efficiency is about $75\%$. The efficiency falls short of a $\sim 90\%$ transfer rate, because the intermediate state decays only into the initial state. This disparity in the intermediate state decay rates was shown to decrease the efficiency of the STIRAP process in \cite{ivanov2005spontaneous}. However, this efficiency can be enhanced by increasing the pulse energy, see Fig. \ref{prob_vs_detp_1} and Fig. \ref{Prob_vs_omgp}.

To better understand the linewidth of the produced 2S atoms of a thermal ensemble of hydrogen atoms, Sec \ref{subsec_thermal_ens}, we examine the efficiency of the STIRAP process as the pulses detuning is varied. Since the Doppler detuning $\Delta_D = \vec{v}\cdot\vec{k}$ of the Ly-$\alpha$ is 5 orders of magnitude higher than that of the microwave pulse, we only sweep the P pulse detuning while keeping the S pulse detuning fixed. We study the behavior of the system for $\Omega^0_s = \Omega^0_p$ and both are equal to multiples of the Rabi frequency $\Omega_0= 4.915\times10^{8}$~rad/sec. In Fig. \ref{prob_vs_detp_1} we see that the range of the detuning $\delta_p$ that produces population in the 2S state increases by increasing the Rabi frequency $\Omega_0$ and has a linewidth that is approximately linear in $\Omega_0$ \cite{vitanov2001coherent} \cite{danileiko1994landau}. This will be important for controlling the width of the velocity distribution of the resulting 2S atoms from a wide thermal ensemble of initial 1S atoms that experiences Doppler detuning in Sec \ref{subsec_thermal_ens}. 

\begin{figure}[h!] 
  \includegraphics[width=1.0\linewidth]{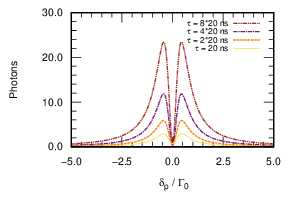} 
\caption{\label{emit_diff_tau} The number of emitted photons as a function of the P pulse detuning $\delta_p$ while keeping the microwave pulse detuning $\delta_s$ fixed for FWHM of the Gaussian pulses taking the values $\sigma = \tau = \{1,2,4,8\}\times20$~ns with $t_0^p = 0.525\tau$. The  Rabi frequencies of the S and P pulses were fixed to $\Omega^0_s = \Omega^0_p = 4.915\times10^8$~rad/sec; thus, the Ly-A pulse energy is changing, Eq. \ref{eq_rabi_prop}.} 
\end{figure}

The number of emitted photons during the decay process from the 2P state to the 1S state is useful in understanding the change in the velocity distribution of the hydrogen atoms. 
High velocity atoms experience high Doppler detunings. Thus, knowing the number of emitted photons at different detunings gives an idea of the amount of recoil experienced by each velocity class of the hydrogen atoms, Sec \ref{subsec_thermal_ens}. In the simulations it was observed that as the detuning goes to zero, the number of photons emitted during the whole STIRAP process approaches its smallest value. In Fig. \ref{emit_diff_tau} it can be seen that there are less photons emitted at zero P pulse detuning $\delta_p$, at fixed $\Omega_p^0$, for longer pulses durations. This happens because the STIRAP efficiency, at high decay rates, $\Gamma_0 > \Omega_0$, \cite{RevModPhys.89.015006,ivanov2005spontaneous}, improves for longer pulses' durations, following the factor $\Omega_0^2 \times \tau / \Gamma $. However, at higher P pulse detuning the maximum number of emitted photons was observed to be higher for longer pulses durations, $\tau$. This is because at high P pulse detuning $\delta_p$, almost no transfer to the 2S state occurs and most of the electrons that make it to the 2P level decay back to the 1S level emitting more photons in the process. Thus, for longer pulse durations, the atoms have enough time to be excited to the 2P level and decay to the 1S level multiple times.

The dependence of the emitted photons on the Rabi frequency $\Omega^0_s = \Omega^0_p$ was observed to be similar to the dependence on the pulse duration. At zero P pulse detuning, the number of emitted photons decreases by increasing the pulse energy, i.e. improving the STIRAP adiabaticity. However, the difference was that the maximum number of emitted photons has a ceiling value that cannot be exceeded, independent of how much the Rabi frequencies are increased. That maximum photon number is proportional to the number of possible excitation cycles to the 2P state that the system can go through over the pulse duration, thus proportional to $\tau \Gamma_0$. Finally, it is worth noting that when the energy in the pulse is kept fixed and the Rabi frequencies are changed with the pulse duration following Eq. \ref{omgp_r}, then neither the transfer efficiency nor the number of emitted photons, at zero pulses' detunings, vary much by changing the pulses' duration. The reasoning for this is that; for the STIRAP efficiency, the dependence on the pulses' duration cancels out from the efficiency factor $\Omega_0^2 \times \tau / \Gamma > 1$ \cite{ivanov2005spontaneous}. We observed that as the pulses' durations increase, the maximum population in the 2P level decreases, but the length of time it remained populated increases, resulting in almost the same number of emitted photons.

Increasing the peak Rabi frequencies of the STIRAP pulses increases the probability of population transfer to the 2S level\cite{kumar2016stimulated} as observed in Fig. \ref{prob_vs_detp_1} and Fig. \ref{Prob_vs_omgp}. For small $\Omega^0_p/ \Omega^0_s$, the final probability to be in the 2S state increases linearly with $(\Omega^0_p)^2$ keeping $\Omega^0_s$ fixed as shown in Fig. \ref{Prob_vs_omgp}. The peak Rabi frequency of the Gaussian P pulse $\Omega^0_p$ can be controlled through the energy in the pulse $\mathcal{E}$, the duration of the pulse $\tau$, or the waist of the Gaussian pulse $w_0$ \cite{svelto2010principles} with: 
\begin{equation}
\label{eq_rabi_prop}
    \Omega^0_p \propto \frac{\sqrt{\mathcal{E}}}{w_0 \sqrt{\tau}}
\end{equation}
For a single atom that is fixed in space, it is irrelevant which of the two parameters $\{w_0,\mathcal{E}\}$ is used to tune $\Omega^0_p$. However, when the decrease of the intensity of the Gaussian P pulse over the spacial extent of the atoms is accounted for in Sec\ref{sec_ly-spatial}, the effects of changing the P pulse energy and its waist will not be equivalent, see Fig \ref{Prob_vs_w0}.

\subsection{STIRAP for a thermal ensemble}
\label{subsec_thermal_ens}
In this section we start with a thermal distribution of hydrogen atoms. This removes one of the idealizations of the previous section and leads to dependencies more relevant for experiments. Atoms are initialized from a Maxwell-Boltzmann distribution with the initial probability that a hydrogen atom has a momentum $\hbar k$ is given by:
\begin{equation}
    \vartheta(k) = \frac{\hbar}{m_H}\sqrt{\frac{m_H}{2 \pi \kappa_B T}} \exp\left(\frac{-\hbar^2 k^2}{2 m_H \kappa_B T}\right)
\end{equation}
where $m_H$ is the mass of the hydrogen atom, $\kappa_B$ is the Boltzmann constant, and $T$ is the temperature.

\subsubsection{The velocity distribution of produced hydrogen atoms}
\label{subsec.thermal_distrib}
\begin{figure}[h!] 
  \includegraphics[width=1.0\linewidth]{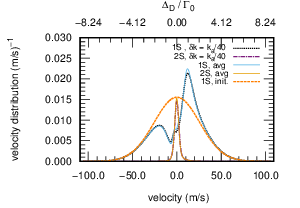} 
\caption{\label{compare_w_v} The velocity distribution of the hydrogen atoms in the 1S and the 2S levels at T=80~mK using the pulses' parameters at the beginning of Sec. \ref{subsec_single_STIRAP}. The solid cyan and orange curves are produced using the averaged approximation, while the dotted black and dot-dashed purple curves are produced using the quantized approach developed in Sec. \ref{subsec.mom_rec} with momentum spacings $\delta k = k_\alpha /40$, with $k_\alpha$ being the wave number of the Ly-$\alpha$ pulse. The dashed orange curve is the initial Gaussian distribution of the hydrogen atoms before STIRAP. The upper x-axis is the Doppler detuning $\Delta_D = \vec{v}\cdot\vec{k}_\alpha$ in units of the decay rate $\Gamma_0$, App. \ref{app.decay_rates} and App. \ref{app.Ham_w_centerofmass}.
} 
\end{figure}

In Fig. \ref{compare_w_v} we initialized the hydrogen atoms at a temperature of $80$~mK and let the system evolve under the master equation Eq. (\ref{master_eq}) using the pulses' parameters discussed at the beginning of Sec. \ref{subsec_single_STIRAP}. We compared the final distribution of the hydrogen atoms in the 1S and the 2S states using the approaches discussed in Sec. \ref{subsec.mom_rec} in Fig. \ref{compare_w_v}. There is a slight difference in the hydrogen atoms distribution in the 1S state between the results from applying averaged approach and the quantized approach. However, the two approaches give similar results for the 2S velocity distribution. We have checked that the result of the quantized approach has converged by comparing the results using two momentum grids with $\delta k = k_\alpha/10$ and $\delta k = k_\alpha/40$.

 There are two key points when studying the velocity distribution of the hydrogen atoms in the 2S level resulting from the STIRAP process. Firstly, when the STIRAP process occurs, the hydrogen atom absorbs a photon with momentum $\hbar k_\alpha$ leading to momentum change of the atom. Secondly, since the atoms at non-zero velocities experience Doppler detuning, only the atoms near the velocity that has zero net detuning $\bar{\Delta} = \Delta' + \vec{v}\cdot\vec{k} = 0$ experience appreciable population transfer from the STIRAP. In Fig. \ref{compare_w_v} the initial hydrogen atoms ensemble was at a temperature $T=80$~mK, the percentage of atoms transferred to the 2S state is approximately $9.6\%$. That percentage is small, compared to the stationary atom $\sim 75\%$, because at $T=80$~mK most of the atoms have a large Doppler detuning, taking most of them off resonance, and little transfer occurs. The width of the velocity distribution of the produced 2S atoms increases as the Rabi frequency increases as expected from the results in Fig. \ref{prob_vs_detp_1}. After STIRAP, the 1S hydrogen atoms velocity distribution is a mixture of the high velocity, highly detuned, 1S atoms of the initial Gaussian distribution, that did not absorb any photons, and the ones that got excited to the 2P state then decayed back to the 1S level. The recoil the atoms experience when absorbing the Ly-$\alpha$ pulse gives them positive velocity kicks, which is seen in the decrease of the number of negative velocity atoms and the increase of positive velocity ones. This suggests that if consecutive pulses are used on the same sample, then sending alternating pulses from opposite directions could reduce the heating effects. 

\begin{figure}[h!] 
  \includegraphics[width=1.0\linewidth]{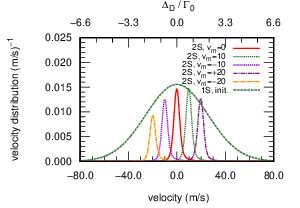} 
\caption{\label{fig_multi_2S_peak} The velocity distribution of the hydrogen atoms in the 2S level at T=80~mK using the pulses' parameters at the beginning of Sec. \ref{subsec_single_STIRAP}. Different detuning for the Ly-$\alpha$ pulse were used to produce hydrogen atoms in the 2S level with the peak velocities $\{-20,-10,0,10,20\}$~m/sec. The dashed green curve is the initial Gaussian distribution of the hydrogen atoms before STIRAP. The upper x-axis is the Doppler detuning $\Delta_D = \vec{v}\cdot\vec{k}_\alpha$ in units of the decay rate $\Gamma_0$, App. \ref{app.decay_rates} and App. \ref{app.Ham_w_centerofmass}} 
\end{figure}

 It is possible to pick the location of the peak of the final velocity of the hydrogen atoms in the 2S level as long as it is inside the thermal distribution by adjusting the pulse detuning $\Delta$ \cite{raizen2014magneto}, see App.\ref{app.Ham_w_centerofmass}. In Fig. \ref{compare_w_v} we adjusted the detuning such that the peak of the hydrogen atoms in the 2S state after receiving the momentum kick would be at zero velocity using:
\begin{equation}
    \Delta' = \delta' = \hbar \frac{ k^2_\alpha}{m_H} \sim 0.27 \Gamma_0
\end{equation}
We show the feasibility of controlling the peak velocity of the produced hydrogen atoms from the STIRAP process in Fig. \ref{fig_multi_2S_peak}. We use an initial ensemble at temperature $T=80$~mK and by adjusting the detuning of the Ly-$\alpha$ pulse we produce hydrogen atoms in the 2S level with peak velocities $v_m =\{-20,-10,0,10,20\}$~m/s. Since the atoms in the 2S level have received a velocity kick $v_k$ from absorbing the Ly-$\alpha$ pulse, the peak of 2S atoms velocity distribution is proportional to the Gaussian Maxwell-Boltzmann distribution evaluated at $v_m - v_k$ instead of just $v_m$. This feature could be experimentally useful if a specific velocity of 2S atoms was desired.

\begin{figure}[h!] 
  \includegraphics[width=1.0\linewidth]{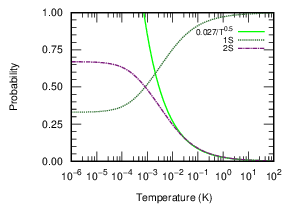} 
\caption{\label{Prob_vs_T} The total probability of the hydrogen atoms being in the 1S and the 2S levels as a function of the initial temperature of the hydrogen atoms ensemble using the pulses' parameters at the beginning of Sec. \ref{subsec_single_STIRAP}. The detuning $\Delta'$ in the Hamiltonian, Eq. (\ref{final_3x3_ham_k}), was set to zero for all temperatures. We also plot the high temperature asymptotic behavior curve $\propto 1/\sqrt{T}$
} 
\end{figure}

In Fig. \ref{Prob_vs_T} we examine how the percentage of the atoms that end up in the different states changes as we change the temperature. We see that at high temperatures, the momentum range that gives a Doppler detuning within the STIRAP transfer domain $\sim [-\Omega_0:\Omega_0]$ becomes small compared to $\kappa_B T$. Thus, the factor $\exp\left(\frac{-\hbar^2 k^2}{2 m_H \kappa_B T}\right)$ becomes almost unity, and the transfer rate dependence on the temperature goes like $\frac{1}{\sqrt{T}}$. For the pulses parameters discussed at the beginning of Sec. \ref{subsec_single_STIRAP}, this dependence was observed for temperatures higher than about $T=10$~ mK. 
For low temperatures, $T<0.1$~mK, the thermal velocities become very small and the Doppler detuning becomes negligible, so one gets the high transfer rate observed for zero detuning in Sec \ref{subsec_single_STIRAP}. For instance, temperatures lower than $T=0.01$~ mK have a small initial velocity spread and approximately 66\% of the atoms' population transfer to the 2S level with an average momentum $\hbar k_\alpha$. The reason why this rate is not exactly the $\sim75\%$ in Fig. \ref{prob_omg_vs_t} for stationary atoms is that here the atoms decaying from the 2P to the 1S level can experience high Doppler detuning due to the decay recoil not canceling out the momentum kick from absorbing the Ly-$\alpha$. From our simulations, we find that increasing the Rabi frequency of the pulses does not change the overall dependence on the temperature; it only shifts the temperature at which the asymptotic $\propto 1/\sqrt{T}$ behavior starts. For instance, at 8 times the Rabi frequency, the transfer ratio becomes about $50\%$ near $T=100$~mK.  We note here that cooling with optical molasses can get the hydrogen atoms to minimum temperatures of only a few milliKelvins \cite{donnan2012proposal}\cite{walsh2024simulated}.

 Although lower-temperature ensembles of hydrogen atoms produce a higher percentage of atoms in the 2S level, they are less tolerant to changes in detuning. For example, this is observed when comparing the percent of atoms that transfer to the 2S level at the two temperatures $T=1$~mK and $T=80$~mK. At zero detuning the lower temperature ensemble transfers about $50\%$ of the atoms to the 2S level, while the ensemble at $T=80$~mK transfers only about $10\%$. However, the lower-temperature ensemble has a width of only about one $\Gamma_0$, while the higher-temperature ensemble has a width almost three times as much. This happens because at higher temperatures the Maxwell-Boltzmann distribution velocity range is larger. Thus, such atoms' Doppler detuning can cancel out the laser detuning, allowing for transitions at a larger laser detuning range.

\subsubsection{Including the spatial profile of the Ly-$\alpha$ pulse}
\label{sec_ly-spatial}

\begin{figure}[h!] 
  \includegraphics[width=1.0\linewidth]{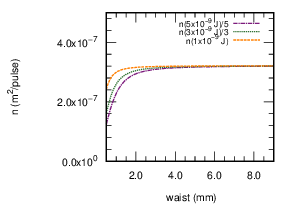} 
\caption{\label{Prob_vs_w0} 
The dependence of the number of hydrogen atoms transferred to the 2S level per pulse on the pulse waist for the Ly-$\alpha$ pulse energies $\{1,3,5\}$~nJ, temperature $T=80$~mK and S pulse Rabi frequency $\Omega^0_s = 4.915\times10^8$~rad/sec. The curves were scaled to show the linear dependence on the energy at large beam waist. 
} 
\end{figure}

In this section, we remove the last idealization of Sec. \ref{subsec_single_STIRAP} and evaluate the number of 2S atoms taking into account the radial profile of the Ly-$\alpha$ beam. Since the Ly-$\alpha$ is typically a Gaussian beam in the radial direction, not all trapped atoms experience STIRAP with the same Rabi frequency $\Omega^0_p$ due to the spatial dependence of the intensity, Eq. (\ref{omgp_r}). However, the S pulse is constant in the radial direction, which results in pulses' configuration different from that of perfect STIRAP and rather a STIRAP-like one. A sample of number density, $\phi$, and length, $L$, would produce, N, hydrogen atoms in the 2S level with:
\begin{eqnarray}
   N &=&  n L \phi =L \phi \int_0^\infty 2\pi r P_{2S}(r) dr 
\end{eqnarray}
where $n$ gives the number of atoms in the 2S level in units of $m^2/pulse$. We introduce $n$ as a cross section like parameter since it is independent of the sample density and length. 
The $P_{2S}(r)$ is the transition probability into the 2S level by the end of the pulse duration, which depends on the specific Rabi frequency at radial distance, $r$, from the center of the path of the pulse
\begin{equation}
\label{omgp_r}
    \Omega^0_p(r) \propto e^{-r^2/w_0^2} \frac{\sqrt{\mathcal{E}}}{w_0 \sqrt{\tau}}
\end{equation}
where the variation in the pulse waist is negligible over the sample size as the Rayleigh length is larger than 6 meters for $w_0>0.5$~mm.  A rough estimate for the parameter $n$ could be obtained by multiplying the probability of transfer at a certain temperature by the beam area $\sim \pi w_0^2$. In Fig. \ref{compare_w_v}, the transfer probability was about $9.6\%$, giving $n\sim 1.2\times10^{-6}$ which is close to the actual value of $\sim0.9\times10^{-6}$ in Fig. \ref{Prob_vs_w0}.


In Fig. \ref{Prob_vs_w0} we show the number of atoms in the 2S level as a function of the pulse waist for different pulse energies $\{1,3,5\}$~nJ. We see that even though the decrease in pulse waist increases $\Omega^0_p$, the number of transferred hydrogen atoms decreases as the waist decreases. This happens because the effective spatial extent of the pulse is of order $w_0$. Thus, decreasing $w_0$ reduces the effective region where efficient STIRAP occurs. However, the large waist behavior could be understood by referring to Fig. \ref{Prob_vs_omgp}, where for small $\Omega^0_p$ and fixed $\Omega^0_s$ the transfer probability is linear in $(\Omega^0_p)^2$. Thus, for large $w_0$, that is, small $\Omega^0_p$, the radial integral of the probability with $\Omega^0_p$ in Eq. (\ref{omgp_r}) turns out to be independent of the waist. The dependence on the pulse energy could be understood in the same manner, where the number of transferred hydrogen atoms is approximately linear in the P pulse's energy at large waist. We investigated the velocity distribution of the 2S hydrogen atoms after including the spatial profile of the Ly-$\alpha$ pulse and got similar results to Fig. \ref{fig_multi_2S_peak} with the same relative peak amplitudes and distribution linewidth. 



\subsection{$1S_c \rightarrow 2S_c$ STIRAP in the presence of magnetic field}
\label{subsec_res_inc_B}

It is possible to transfer population into another hyperfine sublevel of the 2S level, namely the $2S_c$ sublevel. This is achieved using a magnetic field to split the energy levels \cite{martin1995coherent,martin1996coherent,shore1995coherent}. In Sec. \ref{subsec.B_methods} we studied how the magnetic field affects the STIRAP Hamiltonian connecting the states $1S_c$ to $2S_c$. The magnetic field introduces a shift in the energy that increases as the magnetic field increases. We showed that this energy shift could be absorbed in the detuning in Eq. (\ref{eq_B_det}). For a magnetic field of $0.05$~T that shift is about:
\begin{eqnarray}
    2B \frac{g_e e}{\hbar m_e} - E^{1S}(B) &=& 11.12 \Gamma_0 \nonumber \\
    E^{2S}(B) - E^{1S}(B) &=& 7.85 \Gamma_0   
\end{eqnarray}
which shifts the energies of the states off resonance from the Ly-$\alpha$ pulse. A population transfer between the desired levels is achieved by adjusting the pulses' detunings $\{\Delta, \delta\}$.

\begin{figure}[tp]
    \centering
    \subfigure{\includegraphics[width=1\columnwidth]{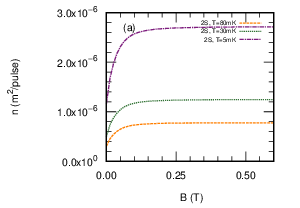} \label{2S_B_vs_detp}}
    \subfigure{\includegraphics[width=1\columnwidth]{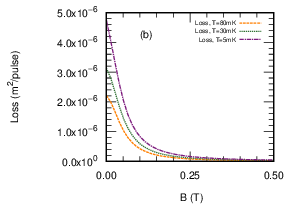} \label{Loss_B_vs_detp}}
    \caption{The number of hydrogen atoms in the $2S_c$ state \ref{2S_B_vs_detp} as well as the number of atoms leaving the subspace $\{1S_c, 2P_a^{\downarrow} , 2S_c\}$ \ref{Loss_B_vs_detp} as a function of the applied magnetic field with zero detunings at temperatures $T=\{5, 30 , 80\}$~mK. The P pulse had a waist $w_0 =1.0$~mm and an energy of $3$~nJ.}
\end{figure}

As discussed in Sec. \ref{subsec.B_methods}, the main effects of introducing a magnetic field on the Hamiltonian would be the factor of $\beta_i$(B) in the off-diagonal terms. For a magnetic field of $0.05$~T these factors are $\beta^2_{1S} = 0.854$ and $\beta^2_{2S} = 0.995$ while for B~$0.15$~T they become $\beta^2_{1S} = 0.974$ and $\beta^2_{2S} = 0.999$. In treating the decay, only a $\beta^2_{1S}$ fraction of the atoms in the 2P state decay into the $1S_c$ state while the remaining fraction $1-\beta^2_{1S}$ decay to the untrappable $1S_a$ state. Thus, a fraction $1-\beta^2_{1S}$ of the atoms leaves the three states subsystem $\{1S_c,2P_a^{\downarrow},2S_c\}$. Since the magnetic fields of interest have the $1S_a$ state far off resonance, there is almost no transfer of population out of the $1S_a$ state. Since the factor $\beta^2_{1S}$ approaches unity as the magnetic field increases, the population leakage out of the subsystem decreases for larger magnetic fields. This is shown in Fig. \ref{Loss_B_vs_detp} where the number of hydrogen atoms lost outside the subsystem is evaluated for different values of the magnetic field. The number of atoms decaying into the $1S_a$ state decreases about 10 times as the magnetic field increases from 0 to 0.15~T. Similarly, the presence of the magnetic field enhances the STIRAP transfer efficiency, as shown in Fig. \ref{2S_B_vs_detp} where the number of atoms moving to the $2S_c$ level more than doubled as the magnetic field increased from 0 to 0.15~T. Magnetic fields bigger than 0.15~T hardly improve the transfer rates, as $\beta^2_{1S}$ becomes almost unity. At lower temperatures the loss in population is higher for the same magnetic fields because the lower Doppler detuning experienced by the atoms at lower temperatures leads to more $1S_c$-$2P_a$ transitions. 


Other than the loss in the population due to decaying with spin flip, the behavior of the system with a magnetic field is similar to that studied in the previous subsection. For instance, when varying the detuning, the transfer probability to the 2S level is the same as that without including a background magnetic field. The only difference being a slight decrease of the transfer rate to the 2S level in the presence of a magnetic field due to the loss of trapped atoms through decaying into the $1S_a$ state. However, as the magnetic field increases, $\beta_{1S}^2$ gets closer to unity and the system behaves like the closed three-state system discussed in the previous subsection.

\section{Conclusion}
\label{sec:Conclusion}
We investigate conditions for the validity of a general STIRAP procedure from the 1S to the 2S levels in the hydrogen atom. We utilize a Ly-$\alpha$ pulse for the 1S-2P coupling and a microwave pulse for the 2P-2S coupling. The stretched states $\{1S_d, 2P_a^{\uparrow}, 2S_d\}$ form a $3\times3$ closed Lambda linkage between which the STIRAP process could be utilized without the need for an external magnetic field. We also showed that by applying a magnetic field, the STIRAP process is also achievable between the states $\{1S_c, 2P_a^{\downarrow}, 2S_c\}$. We found that significant transfer rates through STIRAP occur when the detuning of the P pulse is approximately smaller than the maximum of the Rabi frequency. It was also observed that at fixed microwave pulse's Rabi frequency, the transfer efficiency of STIRAP had a quadratic dependence on the Rabi frequency of the Ly-$\alpha$ pulse for small Rabi frequencies.

We study thermal ensembles of hydrogen atoms and their resulting velocity distribution after the STIRAP process. Atoms at temperatures of a few tens of millikelvins experience Doppler detunings of comparable order of magnitudes to the energies in the system. This Doppler detuning takes the atoms off resonance with the Ly-$\alpha$ pulse and thus affects the efficiency of the STIRAP. The transition probabilities into the 2S level were larger at lower temperatures and proportional to $\propto 1/\sqrt{T}$ for temperatures higher than 100~mK for P pulse Rabi frequency $\Omega^0_p=4.915\times10^{8}$~rad/s. The temperature at which the $1/\sqrt{T}$ dependence begins increases with increasing the Rabi frequency of the Ly-$\alpha$ pulse. We showed that it is possible to control the peak velocity of the resulting hydrogen atoms in the 2S level by changing the Ly-$\alpha$ pulse's detuning. 

We examined, taking into account the spatial Gaussian profile of the Ly-$\alpha$ pulse, the effect of changing the Ly-$\alpha$ pulse waist and energy on the number of 2S hydrogen atoms produced. Generally, increasing the Ly-$\alpha$ pulse energy increases the number of hydrogen atoms undergoing STIRAP. However, the number of atoms transferred through STIRAP increased with the pulse waist for small Ly-$\alpha$ beam waists. It then reaches a plateau where the increased area of the beam is countered by the decrease in the pulse intensity and increasing the waist has little effect. These findings assume that the pulses are not depleted by the atoms in the gas as well as that the trapped hydrogen atoms fill all the spatial extent of the Ly-$\alpha$ pulse.

 Finally, we study the effects of applying an external magnetic field on the STIRAP process. We showed that the STIRAP process could be applied to transfer population between trappable states inside a magnetic field. For the states $\{1S_c,2S_c\}$ increasing the magnetic field keeps the population in the desired trappable states by sending the other states off-resonance with the microwave and Ly-$\alpha$ pulses. The branching ratio of the decay of the $2P_a^{\downarrow}$ into the trappable state $1S_c$ approaches unity as the magnetic field increases. Consequently, loss in hydrogen atoms through decaying into the untrappable $1S_a$ state decreases with increasing magnetic field. Thus, it is possible to control the target states into which the STIRAP transfers the hydrogen atoms.

As discussed in Sec. \ref{subsec_single_STIRAP} we experimented with different possible parameters for the microwave pulse's configuration, and the ones used in the simulations were the optimal ones found. The Ly-$\alpha$ pulse was restricted to a Gaussian pulse for experimental purposes. Further STIRAP optimization techniques such as optimal pulse control and the Stimulated Raman Exact Passage (STIREP) \cite{stefanatos2021optimal,laforgue2022optimal,liu2023optimal} could be future research directions if needed.

\begin{acknowledgments}
This work was supported by the U.S. National Science Foundation under Grant No. 2409162-PHY and NSERC, NRC/TRIUMF, CFI
\end{acknowledgments}

\appendix
\section{Derivation of the Hamiltonian}
This appendix shows the details of evaluating the STIRAP Hamiltonian used in this work, Eq. (\ref{full_Ham}). First, we start by simplifying the expressions for the general matrix element connected through the electric field in Eq. (\ref{Pulse_field}). We apply the Wigner-Eckart theorem\cite{georgi2000lie} to obtain the ratios between the different non-zero matrix elements connecting the hyperfine states of the 2P to those of the 1S and 2S levels. 
\subsection{Stationary atoms STIRAP Hamiltonian}
\label{app.eval_Hammat}
Starting with the wave function:
\begin{equation}
    \left|\psi\right> = c_1(t) e^{-i\varepsilon_1 t /\hbar}\left|1\right> + c_2(t) e^{-i\varepsilon_2 t /\hbar}\left|2\right> + c_3(t) e^{-i\varepsilon_3 t /\hbar}\left|3\right>
\end{equation}
with the states corresponding to the 1S, 2P and 2S states in ascending order. The coupling to the Ly-$\alpha$ and the microwave fields, Eq. (\ref{Pulse_field}), leads to the evolution of the amplitudes in the states as follows:
\begin{eqnarray}
\label{eq.evo_c_full}
    i\dot{c}_1 &=& -e^{i(\omega_{\alpha}-\omega_{12})t} \Omega_p  c_2/2 \nonumber \\
    i\dot{c}_2 &=& -e^{-i(\omega_{\alpha}-\omega_{12})t} \Omega_p^{\ast} c_1/2 - e^{-i(\omega_s - \omega_{32}) t} \Omega_s^* c_3/2  \nonumber \\
    i\dot{c}_3 &=& - e^{i(\omega_s - \omega_{32}) t} \Omega_s c_2/2
\end{eqnarray}
where the Rabi frequencies from the P and S pulses are given by:
\begin{eqnarray}
\Omega_p &= -\frac{e E^0_p}{\hbar \sqrt{2}} \left<1|x -i y|2\right> &= - e E^0_p \left<1|r_-|2\right> / \hbar \nonumber \\
\Omega_s &= - \frac{e E^0_s}{\hbar \sqrt{2}} \left<3|x -i y|2\right> &= - e E^0_s \left<3|r_-|2\right> / \hbar  
\end{eqnarray} 
with transition frequencies $\omega_{12} = \varepsilon_2-\varepsilon_1$ and $\omega_{32} = \varepsilon_2-\varepsilon_3$. The counter propagating terms involving $\Omega_p^{\ast}$ as well as the off-resonant ones involving $E_s^0$ were omitted in the equation for $\dot{c}_1$ and their corresponding ones in the time derivative of $c_2$. Similar terms were omitted in the time derivative of $c_3$ and their corresponding ones in $c_2$. The final step is to redefine $c_2 \rightarrow e^{i\Delta t} c_2$ and $c_3 \rightarrow e^{i\delta t} c_3$, with $\Delta=\omega_{\alpha}-\omega_{12}$ and $\delta=\omega_{\alpha} - \omega_s - \omega_{12} + \omega_{32}$, to get the simplified expressions:
\begin{eqnarray}
    i \dot{c}_1 &=& -\frac{1}{2} \Omega_p c_2 \nonumber \\
    i \dot{c}_2 &=& -\Delta c_2 - \frac{1}{2} \Omega_p^{\ast} c_1 -\frac{1}{2} \Omega_s c_3 \nonumber\\
    i \dot{c}_3 &=& -\delta c_3 - \frac{1}{2} \Omega_s^{\ast} c_2
\end{eqnarray}
This reproduces the Hamiltonian in Eq. \ref{full_Ham} with $\Delta$ and $\delta$ being the single-photon and the two-photon detunings respectively.

\subsection{Including the atom's motion effects on the STIRAP Hamiltonian}
\label{app.Ham_w_centerofmass}

Here we will present the STIRAP Hamiltonian when the center of mass motion is taken into account. For the treatment here, we let the initial hydrogen atom be in the state $\left|i\right> = 1S_d \otimes \left|\vec{k}\right>$, where the first part is the internal state 1S and the second part is the state of the center of mass having momentum $\hbar \vec{k}$. After absorbing the Ly-$\alpha$ pulse, the center of mass would gain a momentum kick of $\hbar \vec{k}_\alpha = \hbar 2 \pi/ (121.6\times10^{-9})$~$kg.m/sec$ or equivalently a velocity shift of $\vec{v}_k \sim 3.25 \hat{k}_\alpha$~m/sec. Since the wavelength of the photon that couples the 2P and the 2S states is orders of magnitude larger than that of the Ly-$\alpha$, we assume here that the 2P and the 2S states have the same momentum of $\hbar (\vec{k} + \vec{k}_\alpha)$. Thus after including the center-of-mass kinetic energy, the Hamiltonian becomes:

\begin{equation}
\label{Ham_w_centerofmass}
    H^k = \hbar \begin{pmatrix}
        \frac{\hbar k^2}{2m_p} & -\frac{\Omega_p}{2} & 0 \\
    -\frac{\Omega_p}{2} & -\Delta + \hbar \frac{k^2 + 2\vec{k}\cdot\vec{k}_\alpha + k^2_\alpha}{2m_p} & -\frac{\Omega_s}{2}  \\
    0 & -\frac{\Omega_s}{2} & -\delta + \hbar \frac{k^2 + 2\vec{k}\cdot\vec{k}_\alpha + k^2_\alpha}{2m_p}
    \end{pmatrix}
\end{equation}
To better understand this Hamiltonian, we shift the energy by subtracting $\hbar^2 k^2 /2m_p$ from the diagonal and identify $2\vec{k}\cdot\vec{k}_\alpha / 2m_p$ as the usual Doppler detuning $\Delta_D = \vec{v}\cdot\vec{k}_\alpha$. Thus we end up with the Hamiltonian:

\begin{equation}
\label{final_3x3_ham_k}
    H = \hbar \begin{pmatrix}
        0 & -\frac{\Omega_p}{2} & 0 \\
    -\frac{\Omega_p}{2} & -\Delta' + \vec{v}\cdot\vec{k}_\alpha  & -\frac{\Omega_s}{2}  \\
    0 & -\frac{\Omega_s}{2} & -\delta' + \vec{v}\cdot\vec{k}_\alpha 
    \end{pmatrix} 
\end{equation}

This is the full Hamiltonian that includes the effect of having a moving center of mass for the hydrogen atom. Note that the factor $\hbar \frac{ k^2_\alpha}{2m_p}$ has a value of approximately an eighth of a $\Gamma_0$ and was absorbed in the definition of the detuning such that:
\begin{eqnarray}
\label{detun_defin}
    \Delta' &=& \Delta -\hbar \frac{ k^2_\alpha}{2m_p} \nonumber\\
    \delta' &=& \delta -\hbar \frac{ k^2_\alpha}{2m_p}
\end{eqnarray}

Thus, the main change in the Hamiltonian would be to include the Doppler detuning in the relevant diagonal terms. 

\section{Decay channels and rates}
\label{app.decay_rates}
The decay rates $\Gamma_n$, from the different 2P states to the possible hyperfine levels of 1S are:

\begin{eqnarray}
\label{eq.decay.rates}
    \Gamma_1 (\left|2 , 2\right> \rightarrow \left| 1 \right>) &=& \Gamma_0 \nonumber \\
    \Gamma_2 (\left|2 , 1\right> \rightarrow \left| 1 \right>) &=& \Gamma_0 / 2 \nonumber \\ 
    \Gamma_3 (\left|2 , 1\right> \rightarrow \left| + \right>) &=& \Gamma_0 / 2 \nonumber \\ 
    \Gamma_4 (\left|2 , 0\right> \rightarrow \left| 1 \right>) &=& 2\Gamma_0 / 3 \nonumber \\ 
    \Gamma_5 (\left|2 , 0\right> \rightarrow \left| + \right>) &=& \Gamma_0 / 6 \nonumber \\ 
    \Gamma_6 (\left|2 , 0\right> \rightarrow \left| -1 \right>) &=& \Gamma_0 / 6 \nonumber \\ 
    \Gamma_7 (\left|1 , 1\right> \rightarrow \left| 1 \right>) &=& \Gamma_0 / 6 \nonumber \\ 
    \Gamma_8 (\left|1 , 1\right> \rightarrow \left| + \right>) &=& \Gamma_0 / 6 \nonumber \\ 
    \Gamma_9 (\left|1 , 1\right> \rightarrow \left| - \right>) &=& 2\Gamma_0 / 3 \nonumber \\ 
    \Gamma_{10} (\left|1 , 0\right> \rightarrow \left| 1 \right>) &=& 2\Gamma_0 / 3 \nonumber \\
    \Gamma_{11} (\left|1 , 0\right> \rightarrow \left| - \right>) &=& \Gamma_0 / 6 \nonumber \\
    \Gamma_{12} (\left|1 , 0\right> \rightarrow \left| -1 \right>) &=& \Gamma_0 / 6 
\end{eqnarray}
where the notation is such that $\Gamma_n (\left| F , m_F \right> \rightarrow \left| f \right>)$ is the decay rate from the 2P state having total angular momentum $F$, with z-component angular momentum $m_F$, to the 1S hyperfine state $\left| f \right>$. The hyperfine 1S states are such that $(\left| 1 \right>, \left| + \right>, \left| -1 \right>)$ form the triplet subspace with decreasing z-component's angular momentum and the $\left| - \right>$ is the singlet state. The value of $\Gamma_0$ is found analytically using the quantization of the electric field \cite{shankar2012principles} and experimentally \cite{PhysRev.148.1} it has the value $\Gamma_0 = 6.27\times10^8 sec^{-1}$.

We do not include the branching ratio to the 2S states through spontaneous photon emission because the decay rate is about 15 orders of magnitude smaller than that from the 2P levels to the 1S levels.  













\newpage 
\bibliographystyle{unsrt} 
\bibliography{bibliography} 

\end{document}